\documentclass[12pt,a4paper,english,fleqn]{article}

\usepackage[sort&compress,round,comma,authoryear]{natbib}
\usepackage[T1]{fontenc}
\usepackage{ae,aecompl}

\usepackage[colorlinks=true,urlcolor=blue,citecolor=blue,linkcolor=red,bookmarks=true]{hyperref}
\usepackage{graphicx}	
\usepackage{float}	
\usepackage{amsmath}	
\usepackage{amssymb}	
\usepackage{xcolor}
\newlength{\lp}
\usepackage[mathscr]{eucal}
\usepackage{amsfonts}
\makeatother
\usepackage{babel}
\makeatother
\begin{document}
\title{The statistical signal for Milgrom's critical acceleration  boundary  being an objective characteristic  of the optical disk.}
\author{D. F. Roscoe (The Open University; D.Roscoe@open.ac.uk)\\ \\ORCID: 0000-0003-3561-7425}
\maketitle
\begin{abstract}
The various successes of Milgrom's MOND have led to suggestions that its critical acceleration parameter $a_0 \approx 1.2\times 10^{-10}\,mtrs/sec^2$ is a  fundamental physical constant in the same category as the gravitational constant (for example), and therefore requiring no further explanation. There is no independent evidence supporting this conjecture.
\\\\	
Motivated by empirical indications of self-similarities on the \emph{exterior} part of the optical disk (the \emph{optical annulus}), we describe a statistical analysis of four large samples of  optical rotation curves and find that quantitative indicators of self-similar dynamics on the optical annulus are irreducibly present in each of the samples. These symmetries lead to the unambiguous identification of a characteristic point, $(R_c,V_c)$, on each annular rotation curve where $R_c \approx f(M,S)$ and $V_c \approx g(M)$ for absolute magnitude $M$ and surface brightness $S$.
\\\\
This opens the door to an investigation of the behaviour of the associated characteristic acceleration $a_c \equiv V_c^2/R_c$  across each sample. The first observation is that since $a_c \approx g^2(M)/f(M,S)$, then $a_c$ is a constant within any given disk, but varies between disks. 
\\\\
Calculation then shows that $a_c$ varies in the approximate range $(1.2\pm0.5)\times 10^{-10}\,mtrs/sec^2$ for each sample. It follows that Milgrom's $a_0$ is effectively identical to $a_c$, and his critical acceleration boundary is actually the characteristic boundary, $R=R_c$, on any given disk. Since $a_c$ varies between galaxies, then so must $a_0$ also. In summary, $a_0$ cannot be a fundamental physical constant.  
\end{abstract}
\section{Introduction}
One of the many problems which hinder our understanding of the underlying
primary processes driving spiral galaxies is the fact that such galaxies
are frequently observed to be ``afflicted'' by one of the following:
\begin{itemize}
\item ongoing interactions with external objects;
\item manifest signatures of such interactions in the near past;
\item internal inhomogeneities generating local perturbations;
\item presence of bars;
\item unusually active central regions,
\end{itemize}
and so on. The net effect of these various phenomena is to considerably
complicate the task of identifying the irreducible physics which is intrinsic to the essential nature of the \emph{disk galaxy.}

Regardless of these sources of noise, there exist various indications of scaling relationship signatures which may, or may not, be disputed, both within spiral galaxies themselves and within the environment surrounding them. For example:
\begin{enumerate}
\item the long-standing recognition (\citet{Danver}, \citet{Kennicutt}) that the
spirality of the spiral arms in disk galaxies can be usefully classified
in terms of the logarithmic spiral, $R=R_{0}\exp(b\theta)$;
\item on the basis of MOND, \citet{Milgrom1983b} predicted that the baryonic Tully-Fisher scaling relationship for disk galaxies  \[V_{flat}^4 = a_0\, G\, M\] for total mass $M$ integrated out to $V_{flat}$ should hold exactly. See \citet{McGaugh2020} for a modern review of the status of this predicton;
\item the idea that, in the local cosmos at least, material in the IGM is distributed in a hierarchical $D\approx 2$ fashion
\[M \sim 4 \pi \, \Sigma_F \, R^2 \]
about any point, for stochastically determined mass surface density $\Sigma_F$. This idea has been around, in general terms, for a very long time. See \citet{Tekhanovich} for a modern review of the status of this claim.
\end{enumerate}
We take these apparently disparate examples as potential indicators of overlooked deeper symmetries within disks.     
\subsection{The optical annulus hypothesis}\label{Annulus}
The topic of how best to express rotation curves in some generic functional
form has received much attention over the years, but all of these
efforts have been focussed on the whole rotation curve, of which the \emph{Universal Rotation Curve}, first mooted by \citet{PS1991} and developed in detail by \citet{PSS}, provides the archetypal example. The complexity of this approach arises from the fact that, in highly idealized terms, a rotation curve can be considered to consist of three distinct zones: 
\begin{itemize}
	\item a complicated central zone defined over the highly active central regions where concentrations of mass in all forms are typically extreme and energetic; 
	\item a generally smooth rising middle zone encompassing relatively quiet gas, dust and stars;
	\item a more-or-less flat and very quiet external zone encompassing mainly gas.
\end{itemize}
We avoid the difficulties of modelling this tripartite structure by the simple expedient of focussing entirely on the generally smooth and rising RC in the \emph{optical annulus} which comprises the relatively quiet middle zone. 
\\\\
Given that it is possible to extract this optical annulus in some objective way, the most simple possibility that has any chance of accommodating the variation between discs for the rotation curve in this optical annulus is:  
\begin{equation}
\frac{V_{rot}}{V_0}\approx \left(\frac{R}{R_0}\right)^\alpha,\,\,\,R_{min}\leq R
\label{eqn1}
\end{equation}
where $R_{min}$, which defines the inner boundary of the optical annulus, is determined by the extraction algorithm of \S\ref{Extraction}. The outer boundary is not formally defined here but, since optical rotation curves rarely extend to the more-or-less flat external zone, it is taken simply as the maximum extent of measurements in the optical disk. Finally, $(R_0,\,V_0)$ is some point on the annular rotation curve - which, normal practice would assume, can be arbitrarily chosen along the curve. 
\\\\
However, in the following where we analyse large samples of annular rotation curves, we find  that a powerful empirical constraint on annular rotation curves exists which removes any freedom of choice for $(R_0,V_0)$. Instead, this constraint imposes a uniquely defined characteristic point $(R_c,V_c)$  on each annular rotation curve, where $\log R_c\approx F(M,S)$ and $\log V_c \approx G(M)$ for absolute magnitude $M$ and surface brightness $S$. 
\\\\
Observing that, in practice, the explicit form of $\log V_c \approx G(M)$ is identical to a typically calibrated Tully-Fisher relation in the relevant pass-band, then $(R_c,V_c)$ can be properly labelled as the \emph{Tully-Fisher point}. Further investigation  then shows that the characteristic acceleration $a_c \equiv V_c^2/R_c$ $\equiv H(M,S)$ varies in the approximate range $(1.2\pm0.5)\times 10^{-10}\,mtrs/sec^2$ for the samples investigated. It follows that Milgrom's $a_0$ is effectively identical to $a_c$ over these samples, and his critical acceleration boundary is actually the Tully-Fisher boundary, $R=R_c$, on any given disk. 
\section{The rotation curves and data reduction}

The optical rotation curve samples considered in this analysis are
those of: 
\begin{enumerate}
\item \citet{MFB} which, after some reduction, consists of 880 useable ORCs, hereafter
referred to as MFB. This sample was used by \citet{PS1995} as the basic resource for the construction of the Universal Rotation Curve in \citet{PSS}. $I$-band photometry used; 
\item \citet{MF} which consists of 1129 useable ORCs, hereafter
referred to as MF. The objects in this sample are significantly more distant than those of MFB.  $I$-band photometry used; 
\item \citet{SC} which consists of 305 useable ORCs, hereafter referred
to as SC. $R$-band photometry used; 
\item A composite sample collected from \citet{DD1997}, \citet{DD1998}, \cite{DD1999} and \citet{DD2000} consisting of 497 useable ORCs, hereafter
referred to as DGHU. $I$-band photometry used. 
\end{enumerate}
As a general comment, the rotation curves of SC and DGHU are much more heavily
sampled than those of MFB and MF. 

\subsection{Unifying the physical scales for each sample}
Each of the four samples came along with its own Tully-Fisher relation calibrated according to the value of $H_0$ assumed, pass-band used and the definition of optical line-width employed. In line with modern estimates, we re-calibrated all TF relations to $H_0=70/km/sec/Mpc$ using the optical line-width definitions supplied by the sample authors - SC's paper was a comparative study of several different optical line-widths and we simply used the one which he considered the most reliable, the model linewidth $V_{2.2}$. Once this was done, Tully-Fisher absolute magnitudes, surface brightnesses and other physical scales were set accordingly.
\subsection{The folding of rotation curves}\label{sub:The-folding-of}
Notwithstanding the existence of highly accurate folding techniques in which the centre of folding (the kinematic centre) and the systematic recession velocity are treated as two free parameters to be determined such that the folded arms are maximally matched (see for example, the methods of \citet{PS1995}, \citet{RoscoeA} or \citet{CHG}), it was very useful to find that the most basic folding method of all (which simply requires a reasonable estimate of $V_{sys}$ at the photometric centre of the object concerned) is entirely sufficient for the purposes of the present analysis. 
\\\\
That said, one very important point, identified by \citet{PS1995} is
that, given a whole set of velocity measurements over a rotation curve
it is necessary to also have quantitative estimates of the absolute
accuracy of each individual velocity measurement available - and then
to use this information as a means of filtering out only the best
individual measurements for the folding process. Broadly speaking,
any individual velocity measurement was \emph{retained} only if its
estimated absolute error $\leq5\%$. For the MFB, MF, SC and DGHU
samples, this requirement led to losses of $35\%$, $25\%$, $46\%$
and $46\%$ respectively of all individual velocity measurements.
The net effect of these data losses meant that many ORCs were then
left with insufficient data points on them to permit a reliable folding.
The overall attrition rate of ORCs lost to the overall analysis via
this process were $2\%$, $5\%$, $8\%$ and $9\%$ respectively. 
\subsection{The objective extraction of the optical annulus}\label{Extraction}
Since the present analysis depends crucially on the extraction of an objectively defined optical annulus, then we describe the details of this extraction process here in the main text.
\\\\ 
The extraction process operates on the basis that if the optical annulus hypothesis of \S\ref{Annulus} is statistically secure in principle, then the properties of rotation curves in the extracted annulus will confirm that statistical security in fact. Otherwise, they will not. Before continuing, there are some relevant definitions: 
\\\\
By the `whole disc` in this context, we mean the complete set of published
velocity data for the whole disc with the exception of any filtered-out
poor-quality individual measurements (cf last section). The optical annulus
data is extracted from the whole disc data $(R_i,V_i)$, $i=1..N$ using a statistical
algorithm based upon the technique of least-squares linear regression (applied to the model $\log V = \log A + \alpha \log R$) for which
technique the following conventional definitions apply: 
\begin{itemize}
\item an observation is reckoned to be \emph{unusual} if the predictor is
unusual, or if the response is unusual; 
\item for a $p$-parameter model, a \emph{predictor} is commonly defined
to be unusual if its leverage $>3p/N$, when there are $N$ observations.
In the present case, we have a two-parameter model so that $p=2$;
\item similarly, the \emph{response} is commonly defined to be unusual if
its standardized residual $>2$.
\end{itemize}
The extraction of the optical annulus and the computation of $(\alpha,\log A)$ over this annulus, for any given folded and
inclination-corrected rotation curve, can now be described using the
following algorithm: 
\begin{enumerate}
\item Assume the model $V_{rot}\approx A R^{\alpha}$
for the rotation velocity in some annular region of the disc, to be
determined;
\item Form an estimate of the parameter-pair $(\alpha,\log A)$ by a linear regression of $\log V$ data on $\log R$ data for the folded
ORC, initially using the data corresponding to the \emph{whole} folded
ORC; 
\item Determine if the \emph{innermost} observation only is an \emph{unusual}
observation in the sense defined above; 
\item If the innermost observation is unusual, then exclude it from the
data-set and keep repeating the process from (2) above on the continually reducing data-set until the innermost observation \emph{ceases }to be unusual.\emph{ }
\item At the stage when the innermost observation ceases to be unusual,
then the computation has finished.

\begin{enumerate}
\item the resulting reduced data set is considered to define the extracted
annular region. For any given disc, the values of $\left(R,V\right)$
at the remaining innermost observation define the radius, $R_{min}$,
of the interior boundary of the extracted annulus and the rotation
velocity, $V_{min}$,  on this interior boundary;
\item the current values of the parameter pair $(\alpha,\log A)$  characterize the extracted annular rotation curve.
\end{enumerate}
\end{enumerate}
If we consider an extracted annular rotation curve to be useable only if it contains a \emph{minimum} of four data points, then it is found that the MFB sample has reduced to 865 objects, the MF sample to 1088 objects, the SC sample to 283 objects and the DGHU sample to 454 objects.  
\subsection{Statistical outliers}
Because of the risk of introducing cut-off bias in this particular analysis, statistical outliers were identified via a two-stage process. So, for example, starting with one of the raw samples produced in \S\ref{Extraction}, MFB0 say, the first stage identified by eye a very small number of the most unambiguously extreme outliers to create a conservatively reduced first-order sample, MFB1 say. Then an aggressive second-stage parametric process
was applied to create a much sharper reduced sample, MFB2 say - this is where the risk of introducing cut-off bias existed. MFB1 was then used as a control for the statistical quality of MFB2. We find that MFB1 is just a noisier version of MFB2 - there is no evidence of a cut-off bias being introduced into MFB2.
\subsubsection*{First stage}
After the extraction process of \S\ref{Extraction}, every object in each of the four samples is characterized by four parameters, $(\alpha,\log A, M, \log S)$ where $(\alpha, \log A)$ are the power-law parameters and $(M,S)$ are the absolute Tully-Fisher magnitude and surface brightness respectively. For each sample, we used three-dimensional visualization software in the four-dimensional parameter space, $(\alpha, \log A,M,\log S)$, to locate putative outliers by eye. 
\\\\
This process was made relatively straightforward by the fact that the vast bulk of objects in each sample reside on a plane in $(\alpha, \log A,M)$ space and, simultaneously, on a plane in $(\alpha,\log A,\log S)$ space. The most extreme of the outliers are then easily identified as objects which exist far out of the planes concerned.
\\\\
In order to minimize the potential effects of subjective judgement, we limited the number of objects judged to be extreme outliers in this first-stage process to be no more than 2\% of the total sample in each case. In practice, this conservative approach ensured that the identified putative outliers were unambiguously detached from the 98+\% bulk of the sample.
\\\\
The net effect of this process was to reduce the sample sizes of (MFB, MF, SC, DGHU) from (865, 1088, 283, 454) objects respectively to (847, 1072, 279, 444) objects respectively. 
\subsubsection*{Second stage}
The archetypal  annular rotation curve is characteristically smoothly rising and tending towards flatness. It therefore follows as a matter of physical reality that all annular rotation curves must be characterized by $\alpha <  1$. Consequently, the second stage data reduction process simply rejects all objects for which $\alpha \geq 1$. The sample sizes of (MFB, MF, SC, DGHU) were then reduced to (808, 995, 260, 432) objects respectively.
\section{The security of the optical annulus hypothesis}
We demonstrate the statistical security of the optical annulus hypothesis in two distinct ways:
\begin{itemize}
	\item Firstly, by considering the behaviour of the associated residuals, we show that the power-law fits to rotation curves over the annular regions are statistically exact;
	\item Secondly, we demonstrate the objective coherence of the extracted annular optical rotation curves by showing that a scaling relationship $F(M, \log V_{min})=0$, directly analogous to the classical Tully-Fisher relation, exists on the interior boundary of the extracted annulus. This implies that the boundary $R=R_{min}$ is an objectively defined boundary between interior processes and exterior processes. 
\end{itemize}
\subsection{Residuals for each ORC sample} \label{Residuals}
\begin{figure}
	\includegraphics[scale=0.8]{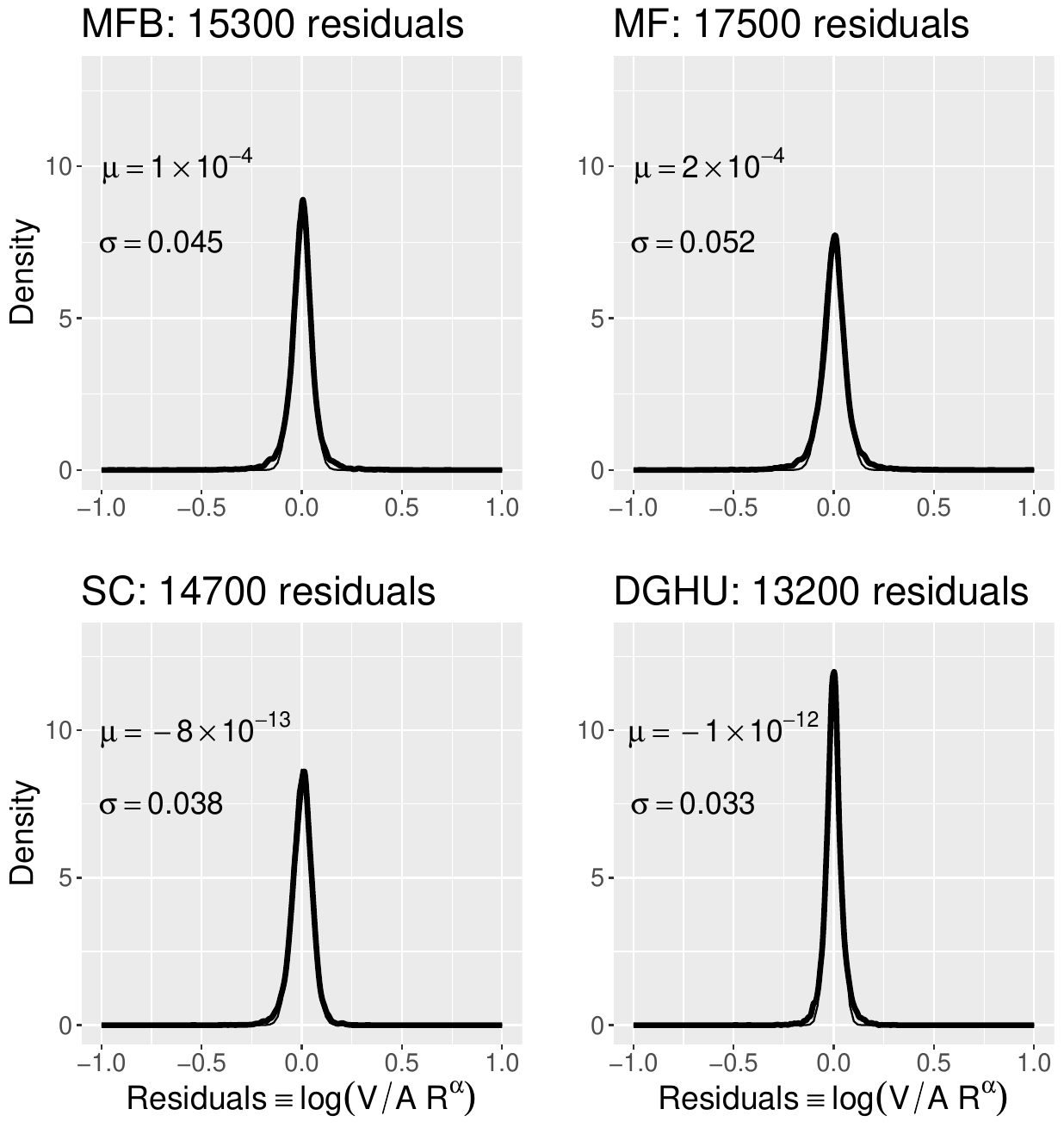}
	\caption{\label{Fig1} Distributions of residuals $e_i \equiv \log \left(V_i/A\,R_i^\alpha \right)$ (thick line) for each of four large ORC samples, overlaid by the best fitting normal distribution $N(\mu,\sigma)$ (fine line) in each case. }
\end{figure}

The basic hypothesis is that rotation curves in the optical annulus extracted according to the algorithm of \S\ref{Extraction} can be modelled by the simple power law:
\begin{equation}
V = A\,R^\alpha \rightarrow \log \left(\frac{V}{A\,R^\alpha}\right) = 0. \label{eqn1A}
\end{equation}
Since any given rotation curve modelled by (\ref{eqn1A}) is characterized by the parameter pair $(\alpha, \log A)$, then a rotation curve with the set of observations $(R_i,V_i),$ $i=1,N$ on it correspondingly has the set of residuals
\[
e_i = \log \left( \frac{V_i}{A\,R_i^\alpha}\right),~~~i = 1,N
\]
associated with it. If a given sample of galaxies contains $M$  individual rotation curves, each containing $N_1, N_2, ...,N_M$ observations respectively, then the whole sample will generate the sequence of residuals:
\[
e_{ij},~~i=1,N_j,~~j=1,M.
\]
The density distribution of the complete set of residuals associated with each of the samples MFB, MF, SC and DGHU is given in Figure \ref{Fig1}. 
\\\\
In each case, the best fitting normal distribution,
$N(\mu,\sigma)$ (fine line), is overlaid onto the residual density plots (thick line) and can barely be distinguished from the corresponding residual density plot. There is no evidence of unexplained residual effects.

\subsection{A scaling law on the interior boundary of the optical annulus} \label{ScalingLaw}
The algorithm described in \S\ref{Extraction} will always produce a result in the
form of an extracted optical annulus. So, the significant questions is: 
\\\\
\emph{ Can
the optical annulus extracted in this way be objectively identified as a physically
coherent distinct sub-component of the whole disc?}
\\\\
In the following, we show that a scaling relation, directly analogous to the classical Tully-Fisher
relation, applies on the interior boundary, $R=R_{min}$, of the optical annulus 
and at similar levels of statistical
significance. This was done by regressing Tully-Fisher absolute magnitudes (calibrated against $H_0=70 km/sec/Mpc$) on  $\log V_{min}$ over each of the four ORC samples to obtain:
\begin{eqnarray}
{\rm MFB:}\,\,\,\,M_{TF} & \approx & (-15.90\pm0.28)+(-2.85\pm0.16) \log V_{min}\,,\nonumber \\
&  & n=808,\,\,\,t_{grad}=-37,\,\,R_{adj}^{2}=63\%\,;\nonumber \\
\nonumber \\
{\rm MF:}\,\,\,\,M_{TF} & \approx & (-16.71\pm0.24)+(-2.66\pm0.14) \log V_{min}\,,\nonumber \\
&  & n=995,\,t_{grad}=-41,\,\,R_{adj}^{2}=62\%\,;\nonumber \\
\label{eqn2}\\
{\rm SC:}\,\,\,\,M_{TF} & \approx & (-13.98\pm0.60)+(-3.37\pm0.28) \log V_{min}\,,\nonumber \\
&  & n=260,\,\,\,\,t_{grad}=-24,\,\,R_{adj}^{2}=68\%\,;\nonumber \\
\nonumber \\
{\rm DGHU:}\,\,\,\,M_{TF} & \approx & (-13.37\,\pm0.62)+(-4.10\pm0.30)\log V_{min}\,,\nonumber \\
&  & n=432, \,\,\,\,t_{grad}=-27,\,\,R_{adj}^{2}=63\%\,,\nonumber 
\end{eqnarray}
where $t_{grad}$ refers to the estimated gradient value. In each case, the values of $(t_{grad},R^2_{adj})$ make it quite obvious that the capacity of $\log V_{min}$ to act as a (noisy) predictor for $M_{TF}$ is established at the level of statistical certainty. In this way, we see that the
interior boundary of the extracted optical annulus has an objective
physical significance and defines, in effect, a transition within the disk between one form
of behaviour and another.
\subsection{Conclusions for the optical annulus}
It has been demonstrated in Figure \ref{Fig1} of \S\ref{Residuals} that the power-law hypothesis (\ref{eqn1A}) for the rotation curve in the extracted optical annulus provides a statistically near-perfect fit to the data over the annulus.
\\\\
Furthermore, the regressions of (\ref{eqn2}) show that a scaling law, directly analogous to that of the classical Tully-Fisher relation and possessing similar levels of statistical significance, exists on the interior boundary of the annulus. This establishes that the extraction process is not arbitrary, but gives a result which has an objective physical significance.
\\\\
These conclusions are powerfully consolidated in the remainder of this analysis.
\section{The signal for $a_0$ in the $(\alpha, \log A)$ diagram}
The wholly unsuspected structure of the $(\alpha,\log A)$ diagram, shown in Figure \ref{Fig2} below, provides the unambiguous signal which reveals that Milgrom's critical acceleration boundary is an objective characteristic of the optical disk, and is the hinge upon which this entire analysis pivots.
\\\\   
After the annular extraction algorithm of \S\ref{Extraction} has been applied to a given sample, then every annular rotation curve in that sample has the two-parameter model form:
\begin{equation}
\log V = \log A + \alpha \log R,    \label{eqn3}
\end{equation}
where the $i$-th rotation curve in the sample is characterized by the parameter pair $(\alpha_i,\log A_i)$.
\\\\
If we now plot $(\alpha_i,\log A_i),$ $i=1..N$ for each of the samples MFB, MF, SC and DGHU, we get the diagrams of Figure \ref{Fig2}. These make it immediately  clear that, in broad terms, the two-parameter model (\ref{eqn3}) is  actually subject to a constraint of the general form $F(\alpha,\log A)=0$, for which the most obvious approximation is 
\begin{equation}
\log A = a + b\, \alpha  \label{eqn4}
\end{equation}
for fixed $(a,b)$, which can be uniquely determined for each sample by a linear regression on $(\alpha_i,\log A_i),$ $i=1..N$ data. Applying this approximation of the constraint to the model (\ref{eqn3}) gives the model annular rotation curve in the one-parameter form:
\[
\left( \log V -a \right) = \alpha \left(\log R + b \right)
\]
from which we can immediately conclude that, for any given sample,
\begin{equation}
\left(\log R_c,\log V_c\right) \equiv \left(-b,a\right)  \label{eqn4A}
\end{equation}
is a characteristic point uniquely associated with that sample. 
After using a linear regression on the $(\alpha_i,\log A_i),$ $i=1..N$ data to estimate $(\log R_c,\log V_c)$ for each sample, and working in units of $kpc$ for $R$, $km/sec$ for $V$ and $mtrs/sec^2$ for $a_c$, we actually find
\\\\
\begin{tabular}{|r|c|c|c|}
	\hline 
	Sample & $ R_c$ & $V_c$ & $a_c\equiv V_c^2/R_c$ \tabularnewline
	\hline 
	MFB & 13.002 & 223.87 & 1.25$\times 10^{-10}$ \tabularnewline
	\hline 
	MF & 14.588 & 222.84 & 1.10$\times10^{-10}$ \tabularnewline
	\hline 
	DGHU & 14.355  & 212.32 & 1.02$\times 10^{-10}$ \tabularnewline
	\hline 
	SC & 14.223 & 231.21 & 1.21$\times 10^{-10}$ \tabularnewline
	\hline 
\end{tabular}
\\\\\\
from which two things are apparent:
\begin{itemize}
	\item the linear approximation (\ref{eqn4}) to the sample plots of Figure \ref{Fig2}  is determined by the large bright objects of each sample;  
	\item within this bright subclass of objects, $a_c \approx 1.2\times 10^{-10}\,mtrs/sec^2$ so that, in effect, Milgrom's critical acceleration boundary is tied to $\left(R_c, V_c \right)$ for these objects in particular.
\end{itemize}
To understand the behaviour of $\left(R_c, V_c \right)$ across the whole range of objects, the simple constraint model (\ref{eqn4}) with (\ref{eqn4A}) must be generalized to account for variations in luminosity properties. 
\begin{figure}[H]
	\includegraphics[scale=0.8]{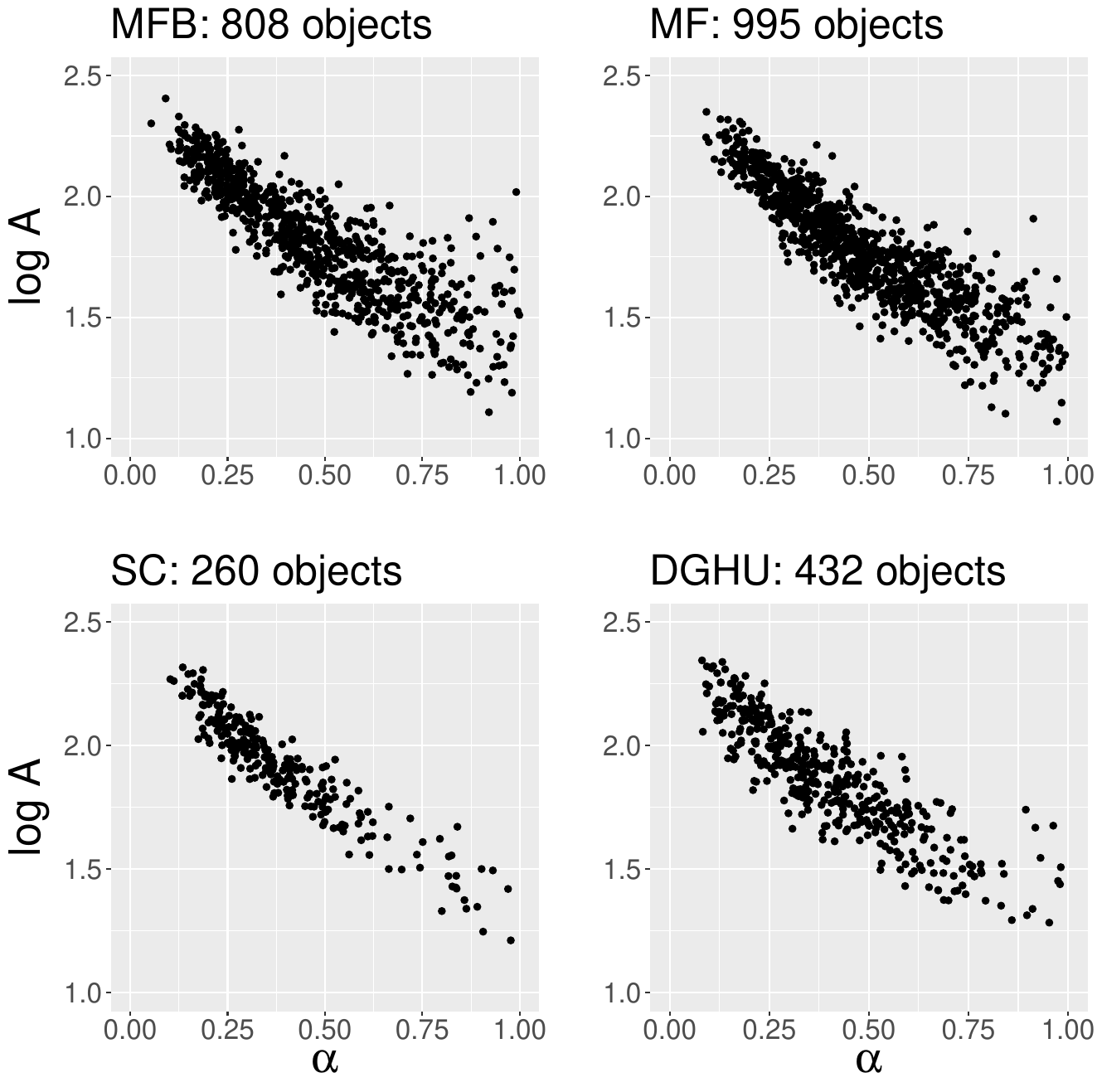}
	\caption{The $(\alpha,\log A)$ scatter plots for each of the four samples of annular rotation curves. } \label{Fig2}
\end{figure}

\section{Generalization: The Tully-Fisher point, $(R_c,V_c)$} \label{TF point}
Using (\ref{eqn4A}), the linear constraint (\ref{eqn4}) can be written as
\begin{equation}
\log A = \log V_c - \alpha \log R_c,   \label{eqn5A}
\end{equation}
for fixed $\left(R_c, V_c \right)$. This is the simplest possible approximation to the plots of Figure \ref{Fig2}, and effectively represents the signal for Milgrom's critical acceleration boundary on the subclass of large bright objects.
\\\\
However, whereas the simple model (\ref{eqn5A}) typically accounts for $<75\%$ of the variation in $\log A$ data in any of the plots of Figure \ref{Fig2}, exploration shows that its generalized form
\begin{eqnarray}
\log A &=& \log V_c - \alpha \log R_c, \nonumber \\
\log V_c &=& a_{0}+a_{1}M, \label{eqn6A} \\ 
\log R_c &=& b_{0}+b_{1}M+b_{2}\log S,\nonumber
\end{eqnarray}
for Tully-Fisher magnitude $M$ and surface brightness $S$,
accounts for up to $96\%$ of the variation in $\log A$ data across all four samples, and is comprehensive - all significant predictors are included. For example, whereas the product $\alpha \log S$ is found to be a significant predictor for $\log A$, the simple term $\log S$ is not, hence its absence from the $\log V_c$ component above. In the following, we illustrate the details of how this generalized model resolves the data of the MFB sample.
\subsection{The Tully-Fisher point illustrated for the MFB sample}

After data reduction, the MFB sample contains 808 useable ORCs. According to (\ref{eqn6A}), the significant predictors for $\log A$ are $(M, \alpha, \alpha M, \alpha \log S)$. A least-squares linear regression then gives a model for $\log A$ defined by the table:
\\\\
~~~~~~~~~~~~~~~~~~%
\begin{tabular}{|l|ccccc|c|}
	\hline 
	Coeffs & $a_{0}$ & $a_{1}$ & $b_0$ & $b_1$ & $b_2$ & $R_{adj}^{2}$\tabularnewline
	\hline 
	Estimate & -0.805 & -0.140 & -3.052 & -0.229 & -0.527 & $90\%$\tabularnewline
	Std Error & 0.129 & 0.006 & 0.286 & 0.011 & 0.022 & \tabularnewline
	$t$-statistic & -6 & -24 & 14 & 21 & 24 & \tabularnewline
	\hline 
\end{tabular}\\
\\
\\
Consequently, for the model (\ref{eqn6A}) we immediately get
\begin{eqnarray}
\log V_c&=&-(0.81\pm0.26)-(0.14\pm0.01) M, \label{eqn7} \\
\log R_c&=& -(3.05\pm0.58)-(0.23\pm 0.02) M-(0.53\pm 0.04)\log S. \nonumber
\end{eqnarray}
The first of these two relations gives, directly 
\begin{equation}
M = -5.79 - 7.14\log V_c \label{eqn5}
\end{equation}
which is a typical $I$-band Tully-Fisher calibration. In other words, the constraint (\ref{eqn6A}) 
recovers the classical Tully-Fisher relation, but with characteristic velocity $V_c$ (rather than the asymptotic velocity), and defines the radial boundary $R_c$ at which $V_c$ is measured. For this reason, we designate  the characteristic point $(R_c,V_c)$ as the \emph{Tully-Fisher point} and $R_c$ as the \emph{Tully-Fisher boundary}. The corresponding analyses for the remaining three samples are similar and are given in appendix \S\ref{AppA}.
\subsection{Acceleration $a_c\equiv V_c^2/R_c$ is a non-trivial function of $M$ and $S$} \label{Accn}
The observation that Milgrom's $a_0$ cannot be a fundamental constant of nature depends critically upon the prior observation that $a_c = H(M,S)$ for non-trivial $H(M,S)$, at the level of statistical certainty. 
\\\\
If we take (\ref{eqn7}) at face value, then the absolute absence of $\log S$ as a predictor in the expression for $\log V_c$ means that $\log V_c$ and $\log R_c$ are linearly independent as a matter of mathematical fact, and this implies directly that 
\begin{equation}
\log a_c \equiv 2 \log V_c-\log R_c \equiv \log 
H(M,S) \label{eqn7B}
\end{equation}
for non-trivial $H(M,S)$. Consequently, to demonstrate that this is indeed the case it is sufficient to show that the omission of $\log S$ from the expression for $\log V_c$ in the model (\ref{eqn7}) is statistically secure.
\\\\
The demonstration is straightforward: we simply compute the regression model
\begin{eqnarray}
\log A &=& \log V_c - \alpha \log R_c, \nonumber \\
\log V_c &=& a_{0}+a_{1}M + a_2 \log S, \label{eqn7A} \\ 
\log R_c &=& b_0+b_1M+b_2\log S\nonumber
\end{eqnarray}   
and observe the status of $\log S$ as a predictor for $\log A$. The linear regression gives the table:
\\\\
\begin{tabular}{|l|cccccc|c|}
	\hline 
	Coeffs & $a_{0}$ & $a_{1}$ &$a_2$ &$b_0$ & $b_1$ & $b_2$ & $R_{adj}^{2}$\tabularnewline
	\hline 
	Estimate & -0.861 & -0.148 & -0.058&-3.147 & -0.242 & -0.624 & $90\%$\tabularnewline
	Std Error & 0.132 & 0.007 & 0.031&0.226 & 0.013 & 0.056 & \tabularnewline
	$t$-statistic & -7 & -20 &-2 &14 & 19 & 11 & \tabularnewline
	\hline 
\end{tabular}
\\\\\\
from which it is immediate that the only significant predictors for $\log A$ are  $(M,\alpha,\alpha M, \alpha \log S)$. There is absolutely no evidence that $\log S$ belongs in this list. Consequently, model (\ref{eqn7}) is indeed secure. It follows by (\ref{eqn7B}) that $a_c = H(M,S)$ for non-trivial $H(M,S)$ and so $a_c$ varies between galaxies.
\section{Accelerations on the Tully-Fisher boundary}
We consider three ways to estimate the distribution of characteristic accelerations $a_c$ on the Tully-Fisher boundary for any given sample: the obvious direct method which leads to a highly attenuated signal; an indirect method based upon a smoothing process which gives a significantly sharper signal; and, finally, a variation of the indirect method which sharpens the signal still further. This latter variation of the indirect method allows the conclusion that $a_c \approx(1.2 \pm 0.5) \times 10^{-10}$$mtrs/sec^2$. We consider each approach in turn. 
\subsection{The direct method + bootstrap estimates}
The obvious way forward is to compute 
$a_c \equiv V_c^2/R_c$
directly for each object using the definitions of $(R_c,V_c)$ for each sample given at (\ref{eqn7}), (\ref{eqn9B}), (\ref{eqn9C}) and (\ref{eqn9D}) respectively, and then to plot the density distributions of $a_c$.
The results of this process are shown as the \emph{thin black} curves in Figure \ref{Fig4} and, notwithstanding its heavy attenuation (almost certainly arising from the noisiness of the photometric data), the signal $a_c \approx 1.2 \times 10^{-10}$$mtrs/sec^2$ is clearly indicated. If we apply a simple bootstrap to the thin black-curve data we find, for the mean values of $a_c$, the estimates: 
\begin{eqnarray}
{\rm MFB}:\,\,\mu(a_c) &=& (1.25 \pm 0.06) \times 10^{-10}mtrs/sec^2; \nonumber \\
{\rm MF}:\,\,\mu(a_c) &=& (1.10 \pm 0.04) \times 10^{-10}mtrs/sec^2; \nonumber \\
{\rm SC}:\,\,\mu(a_c) &=& (1.21 \pm 0.08) \times 10^{-10}mtrs/sec^2; \nonumber \\
{\rm DGHU}:\,\,\mu(a_c) &=& (1.02 \pm 0.06) \times 10^{-10}mtrs/sec^2. \nonumber 
\end{eqnarray}
\subsection{The photometric smoothing method} \label{ModellingMethod}
\begin{figure}[H]
	\includegraphics[scale=0.80]{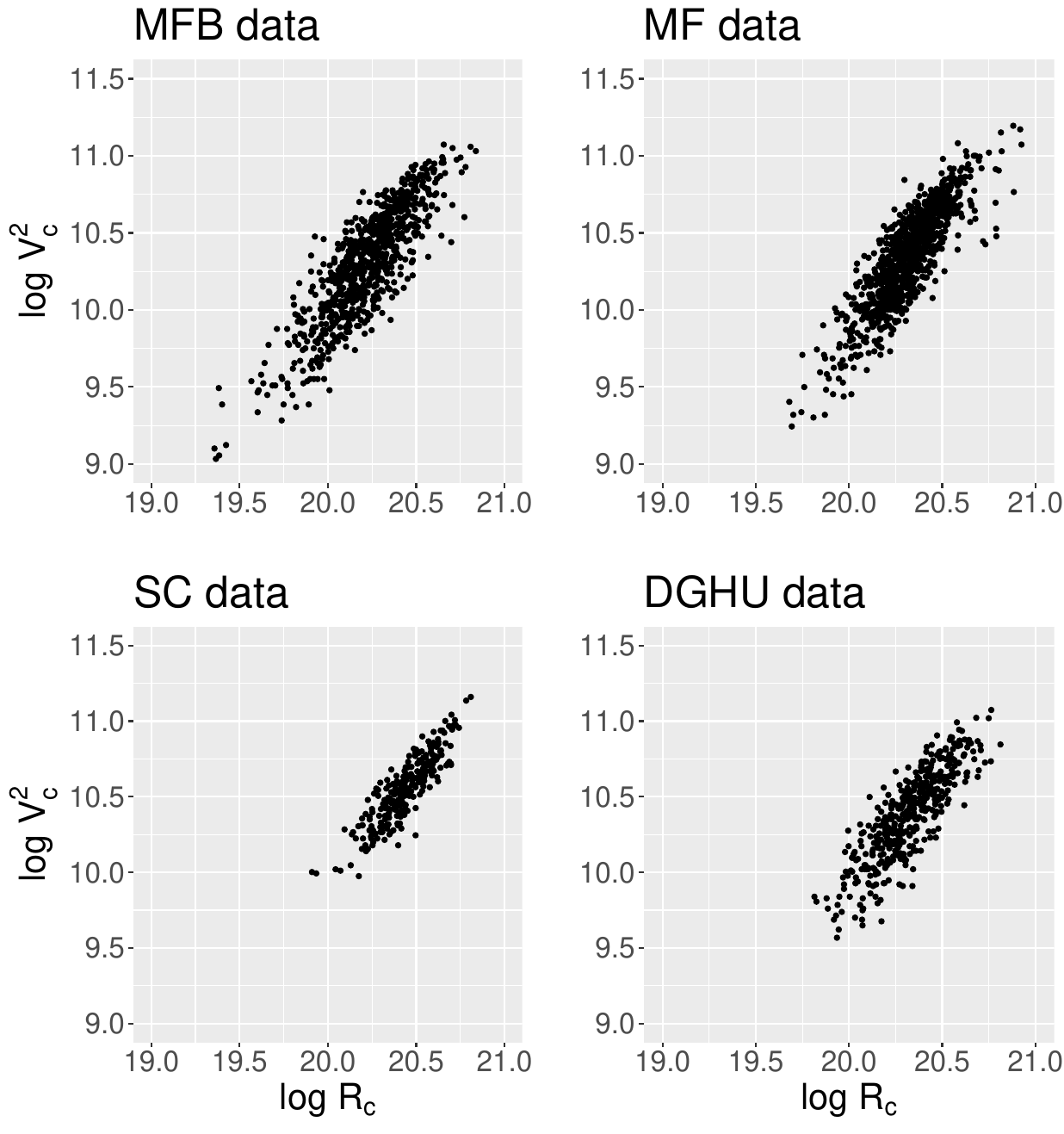}
	\caption{Plots of $\log V_c^2$ against $\log R_c$. Units are mtrs/sec and mtrs. Note the very strong functional dependence between the two variables. } \label{Fig3} 
\end{figure}

The advantage of the smoothing process to be described is simply that in place of estimates of $a_c$ being distorted by two competing sources of noise (a noisy $\log V_c^2$ and a noisy $\log R_c$), there is only the noisy $\log R_c$ to contend with. Consequently, the signal attenuation problem is significantly ameliorated.
\\\\
The smoothing process is suggested by Figure \ref{Fig3}. 
Notwithstanding the fact that $\log V_c^2$ and $\log R_c$ are obviously highly correlated through the photometric modelling process, it is still surprising to find in Figure \ref{Fig3} that the assumption of a simple linear model connecting $\log V_c^2$ and $\log R_c$ is perfectly reasonable. So we have for each of the four samples:
\[
\log V_c^2 \approx \left[\log V_c^2 \right]_{model} \equiv  A_0 \log R_c + A_1 
\] 
from which we can write
\[
\log \left( \frac{V_c^2}{R_c}\right) \approx \left[\log \left( \frac{V_c^2}{R_c}\right)\right]_{model} \equiv P \equiv (A_0-1)\log R_c + A_1.
\]
It is clear that the estimate $a_c \approx a_c(model) \equiv 10^P$ is affected only by $\log R_c$ noise. 
\\\\
So, after using a linear regression to estimate $(A_0,A_1)$ for each of the distributions of Figure \ref{Fig3}, and then using (\ref{eqn7}), (\ref{eqn9B}), (\ref{eqn9C}) and (\ref{eqn9D}) respectively to estimate $R_c$ for each object, we can plot the density distributions of $a_c(model) \equiv10^P$ for each of the four samples. The results are shown as the thick \emph{grey} curves in Figure \ref{Fig4}, where the considerably sharpened signal provides the estimate $a_c \approx (1.2 \pm 0.7)\times 10^{-10}$$mtrs/sec^2$ for the range of variation in $a_c$ driven by the variation in luminosities within each sample.
\subsection{The dynamical smoothing method} \label{Enhanced}
\begin{figure}[H]
\includegraphics[scale=0.80]{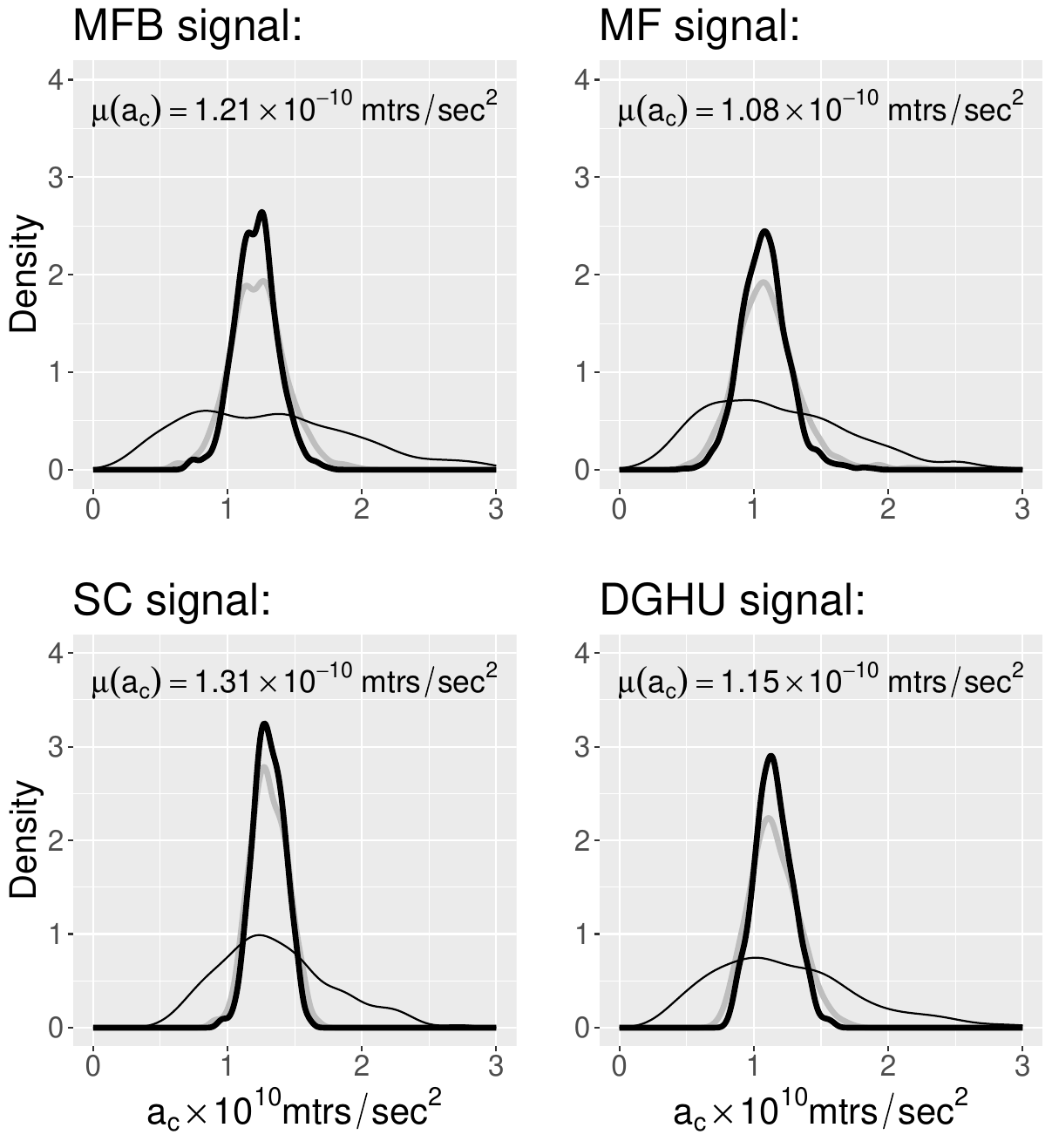}
\caption{Density plots for estimates of $a_c$ on the Tully-Fisher boundary, $R=R_c$. The thin black curve arises from using direct photometric estimates of $(R_c,V_c)$, the thick grey curve arises from the photometric smoothing method using purely photometric estimates of $(R_c,V_c)$  whilst the thick black curve arises from the dynamical smoothing method which chooses each $(R_c,V_c)$ estimate to be that point on the model rotation curve $V=A R^\alpha$ which is closest to the photometric estimate. The quoted values for $\mu(a_c)$ refer to the thick black curves. } \label{Fig4}
\end{figure}

Each rotation curve in the optical annulus is characterized by the four parameters $(\alpha, \log A)$ and $( M, \log S)$. Whilst in an ideal world, any estimate of the characteristic point $(R_c,V_c)$ should lie exactly on its model rotation curve $V = A R^\alpha$, in practice, photometric estimates based on the models (\ref{eqn7}) and similar, will certainly \emph{not} conform to this ideal. So, rather than use the photometric estimate of any given characteristic point, the dynamical smoothing method uses the point on the model rotation curve, $(R^*_c,V^*_c)$ say, which is closest in some sense to the photometric estimate, $(R_c^{**},V_c^{**})$ say.
\\\\
In practice, we choose $\log R^*_c$ as the $\log R$ value which minimizes
\[
\Phi \equiv \left(\log R - \log R_c^{**}\right)^2 + \left(\log A +\alpha \log R - \log V_c^{**} \right)^2,
\]
and then set $V^*_c = A \left(R^*_c\right)^\alpha$. Having generated $(R_c^*,V_c^*)$ for the whole sample in this dynamical way, we apply the basic smoothing  method of \S\ref{ModellingMethod} to these dynamical estimates in place of the purely photometric estimates  $(R^{**}_c,V^{**}_c)$.
\\\\
The results are shown as the thick \emph{black} curves in Figure \ref{Fig4}, where the signal is sharpened still further to provide the estimate $a_c \approx (1.2 \pm 0.5)\times 10^{-10}$$mtrs/sec^2$ for the range of variation in $a_c$ driven by the variation in luminosities within each sample.
\subsection{Consequences for Milgrom's $a_0$}
It was shown in \S \ref{Accn} that $a_c = H(M,S)$ for non-trivial $H(M,S)$ and in \S\ref{Enhanced} that $a_0$ and $a_c$ are numerically identical. The most straightforward conclusion is that $a_0 \equiv a_c$ in which case $a_0 = H(M,S)$ also. Consequently, $a_0$ varies between objects according to their luminosity properties. It cannot be a fundamental physical constant.

\section{Summary and conclusions}
\subsection{Summary}
The optical annulus hypothesis of \S\ref{Annulus}, that on some exterior annulus of the optical disk (defined algorithmically in \S\ref{Extraction}) rotation curves follow a simple power law, $V=AR^\alpha$, was found to be confirmed at the level of statistical certainty by the residual plots of Figure \ref{Fig1}. 
\\\\
Subsequently, the class of all such annular rotation curves were found to be powerfully constrained by a condition of the general form $F(\alpha,\log A)=0$ exemplified by the plots of Figure \ref{Fig2}. A detailed consideration of this constraint quickly shows that it is, in fact, a signal to the effect that Milgrom's critical acceleration boundary is an objective characteristic of the optical disk.
\\\\
In detail, we were led to the existence of a characteristic rotation curve point, $(R_c,V_c)$, where $\log R_c\approx F(M,S)$ and $\log V_c \approx G(M)$ for absolute magnitude $M$ and surface brightness $S$, on each annular rotation curve. Observing that $\log V_c \approx G(M)$ is identical to a typically calibrated Tully-Fisher relation in the relevant pass-band (see (\ref{eqn5})), then $(R_c,V_c)$ can properly be referred to as the \emph{Tully-Fisher point}, and $R=R_c$ as the Tully-Fisher boundary. The identification of the Tully-Fisher point then opened the door to a study of the behaviour of the associated characteristic acceleration, $a_c \equiv V_c^2/R_c$. 
\\\\
It was subsequently shown (Figure \ref{Fig4}) that $a_c$ varies with $M$ and $S$ in the approximate range $(1.2\pm 0.5)\times 10^{-10}$ $mtrs/sec^2$ over the samples analysed confirming that Milgrom's $a_0$ can be identified with $a_c$ and that Milgrom's critical acceleration boundary is, in fact, the characteristic Tully-Fisher boundary, $R=R_c$.
\subsection{Conclusions}
There are three obvious consequences of significance for Milgrom's MOND: 
\begin{itemize}
\item Firstly, the identification of Milgrom's critical acceleration boundary as the characteristic Tully-Fisher boundary, $R=R_c$, provides a key component of the MOND prescription with an objective physical basis that was previously absent.
\item 
Secondly, the observation that the explicit form of $\log V_c \approx G(M)$ for any of the samples is identical to a typically calibrated Tully-Fisher relation in the relevant pass-band has ramifications: specifically, in the classical Tully-Fisher relation, the characteristic rotation velocity is taken to be $V_{flat}$ whilst the mass proxy is taken to be integrated out to $V_{flat}$. By contrast, here the characteristic rotation velocity is $V_c$ on the Tully-Fisher boundary, $R=R_c$, so that in any scaling relationship between rotation velocity and mass, the mass will be that contained within $R=R_c$.
\item Thirdly, the fact that $a_c \equiv V_c^2/R_c$ is a non-trivial function of mass and mass-density proxies means that it varies systematically between galaxy disks. Consequently, the identification $a_0 \equiv a_c$ means that $a_0$ also varies in this way - it cannot be a fundamental constant of physics.  
\end{itemize} 
These points raise new questions and interpretations.
For example, it becomes natural to consider the possibility that the objective Tully-Fisher boundary $R=R_c$ defines, de facto, the galactic boundary. In this case, the statement  $a_c\approx (1.2\pm 0.5)$$\times 10^{-10} mtrs/sec^2$ becomes the statement of a dynamical boundary condition, common to all objects in the four samples. The question then becomes: \emph{What does this dynamical boundary condition secure for the systems in which it is satisfied?} to which an obvious answer might be \emph{Dynamical stability against the external environment}, to which a reasonable response might be: \emph{What is the external environment?}
\\\\
In other words, the results of the present analysis have the potential to completely transform the debate around Milgrom's MOND.

\appendix

\section {The samples of MF, DGHU and SC} \label{AppA}

\subsection*{The sample of MF}
After data reduction, the MF sample contains 995 useable ORCs. According to (\ref{eqn6A}), the significant predictors for $\log A$ are $(M, \alpha, \alpha M, \alpha \log S)$. A least-squares linear regression then gives a model for $\log A$ defined by the table:
\\\\
~~~~~~~~~~~~~~~~~~~~%
\begin{tabular}{|l|ccccc|c|}
	\hline 
	Coeffs & $a_{0}$ & $a_{1}$ & $b_0$ & $b_1$ & $b_1$ & $R_{adj}^{2}$\tabularnewline
	\hline 
	Estimate & -0.970 & -0.145 & -2.646 & -0.203 & -0.445 & $92\%$\tabularnewline
	Std Error & 0.128 & 0.006 & 0.209 & 0.010 & 0.020 & \tabularnewline
	$t$-statistic & -8 & -25 & 13 & 21 & 23 & \tabularnewline
	\hline 
\end{tabular}\\
\\
\\
so that for (\ref{eqn6A}) we immediately get
\begin{eqnarray}
\log V_c &=& -0.97-0.15 M, \label{eqn9B} \\
\log R_c &=&-2.65-0.20 M-0.44\log S.\nonumber
\end{eqnarray}
The first of these two relations gives, directly
\[
M= -6.47-6.67 \log V_c 
\]
which is a typical $I$-band TF calibration.

\subsubsection*{The sample of DGHU}
After data reduction, the DGHU sample contains 432 useable ORCs. According to (\ref{eqn6A}), the significant predictors for $\log A$ are $(M, \alpha, \alpha M, \alpha \log S)$. A least-squares linear regression then gives a model for $\log A$ defined by the table:\\\\
~~~~~~~~~~~~~~~~~~~%
\begin{tabular}{|l|ccccc|c|}
	\hline 
	Coeffs & $a_{0}$ & $a_{1}$ & $b_0$ & $b_1$ & $b_2$ & $R_{adj}^{2}$\tabularnewline
	\hline 
	Estimate & -0.748 & -0.136 & -2.577 & -0.198 & -0.431 & $93\%$\tabularnewline
	Std Error & 0.142 & 0.006 & 0.288 & 0.014 & 0.028 & \tabularnewline
	$t$-statistic & -5 & -21 & 9 & 14 & 15 & \tabularnewline
	\hline 
\end{tabular}\\
\\
\\
so that for (\ref{eqn6A}) we immediately get
\begin{eqnarray}
\log V_c &=&-0.75-0.14 M, \label{eqn9C} \\
\log R_c &=&-2.58-0.20 M-0.43\log S. \nonumber
\end{eqnarray}
The first of these two relations gives, directly
\[
M= -5.36-7.14\log V_c
\]
which is a typical $I$-band TF calibration.

\subsubsection*{The sample of SC using $V_{2.2}$}
After data reduction, the SC sample contains 260 useable ORCs. According to (\ref{eqn6A}), the significant predictors for $\log A$ are $(M, \alpha, \alpha M, \alpha \log S)$. A least-squares linear regression then gives a model for $\log A$ defined by the table:
\\\\
~~~~~~~~~~~~~~~~~~~%
\begin{tabular}{|l|ccccc|c|}
	\hline 
	Coeffs & $a_{0}$ & $a_{1}$ & $b_0$ & $b_1$ & $b_2$ & $R_{adj}^{2}$\tabularnewline
	\hline 
	Estimate & -0.869 & -0.150 & -2.767 & -0.203 & -0.339 & $97\%$\tabularnewline
	Std Error & 0.160 & 0.008 & 0.325 & 0.015 & 0.025 & \tabularnewline
	$t$-statistic & -5 & -20 & 9 & 13 & 14 & \tabularnewline
	\hline 
\end{tabular}\\
\\
\\
so that for (\ref{eqn6A}) we immediately get
\begin{eqnarray}
\log V_c &=&-0.87-0.15 M,\label{eqn9D} \\
\log R_c &=& -2.77-0.20 M-0.34\log S.\nonumber
\end{eqnarray}
The first of these two relations gives, directly
\[
M_{TF}= -5.79 -6.67\log_{10}V_{0}
\]
which is a typical $R$-band TF calibration.

\end{document}